\documentclass[review]{elsarticle}
\usepackage{graphicx,color}
\usepackage{amsmath, amssymb, bm}
\numberwithin{equation}{section}
\newcommand{\lambdabar}{\lambda\kern -0.5em\raise 0.5ex \hbox{--}}

\journal{Journal of \LaTeX\ Templates}

\begin{document}

\begin{frontmatter}

\title{An Approach to Yukawa's Elementary Domain Based on AdS$_5$ Spacetime}


\author[mymainaddress]{Kenichi Aouda}
\author[mymainaddress]{Naohiro Kanda}
\author[mymainaddress]{Shigefumi Naka\corref{mycorrespondingauthor}}
\cortext[mycorrespondingauthor]{Corresponding author}
\ead{naka@phys.cst.nihon-u.ac.jp}
\author[mysecondaryaddress]{Haruki Toyoda}
\address[mymainaddress]{Department of Physics, College of Science and Technology, Nihon 
 University, Tokyo 101-8308, Japan}
\address[mysecondaryaddress]{Junior College, Funabashi Campus, Nihon University, Chiba 
 274-8501, Japan}

\begin{abstract}
The field equations of elementary domains proposed by Yukawa in 1968 are studied from the viewpoint of particle embedded $\mbox{AdS}_5$ spacetime with warp factor. The fifth dimension in $\mbox{AdS}_5$ is known to produce the branes associated with the energy hierarchy as a effect of the warp factor. The particles embedded in this spacetime behave as those in an infinite square well potential due to the boundary conditions in the fifth dimension. Then the superposition of the fields for those particles leads to a Yukawa's domain type of difference equation under some conditions. As a new perspective on AdS spacetime, we investigate the mechanisms that lie between the domain type of field equations in this sense and the spacetime with a compact extra dimension in detail.
\end{abstract}

\end{frontmatter}

\section{Introduction}

The anti-de Sitter space brought various interesting points of view in the modern elementary particle physics since the appearance of the AdS/CFT correspondence\,\cite{Maldacena,Polyakov,Witten}. In particular, the fifth dimension in $\mbox{AdS}_5$ spacetime with the warp factor (RS model) plays a significant role to understand the energy hierarchy of respective branes distinguished by the fifth coordinate\,\cite{Randall-Sundrum-1,Randall-Sundrum-2}; in this sense, the high and low energy ends of the fifth dimension are called respectively as UV brane and IR brane. 

From a quantum mechanical point of view, the fifth dimension gives another interesting aspect for a field in the IR brane.  The fifth coordinate $0\leq y\leq L$ of a particle embedded in AdS${}_5$ spacetime cannot run off the both ends because of the boundary conditions like the infinite square well potential problem. As a result, the Green function in the bulk of AdS${}_5$ spacetime shows a periodic structure with respect to the fifth coordinate\,\cite{Box1,Box2}. Then, if the fifth dimension is inclined toward time direction, the $y$ periodicity of the Green function will be transcribed on a time variable.  When we describe the propagation of the field on the IR brane by this Green function, the timelike periodicity of the Green function is expected to cause a difference type of field equation on the IR brane. Indeed in \cite{Proc}, we outlined the story that a one-dimensional compact extra space, the fifth dimension, attached to $M_4$ spacetime causes a domain type of field equation in $M_4$ due to the periodic structure of the Green function.

Historically, in 1968, a similar type of field equation has been proposed by Katayama-Yukawa-Umemura in their theory of the elementary domain\cite{domain-1,domain-2} as part of non-local field theories. Back in 1949, Yukawa proposed a bi-local field theory\,\cite{Yukawa-1,Yukawa-2} as the start of his works on the non-local field theories. Nowadays, the bi-local field theories are understood from a view point of relativistic two-body problems\,\cite{Takabayasi,Suppl67}; and recently, a new aspect of the bi-local fields was added from a viewpoint of the composite fields\,\cite{AdS-dual-O(N),Collective Fields-1,Collective Fields-2,AdS-BL} in the higher-spin gravity theories based on AdS spacetime.

In those days, however, Yukawa did not get the satisfaction out of the bi-local field theories, and he proposed a more drastic field equation, a difference type of field equation, such that
\begin{align}
e^{\lambda\cdot\partial}\Psi=e^{-\frac{i}{\hbar}\hat{S}_\lambda}\Psi\,,~~~\left(\, \lambda_\mu=\sum_\alpha\lambda_\alpha e^{(\alpha)}_\mu \right)\,,  \label{domain}
\end{align} 
where $\{e^{(\alpha)}_\mu\}$ is a vierbein, and $\hat{S}_{\lambda}$ is an operator depending on the model construction. Yukawa considered this type of field equation is a natural landing in the non-local field theories, although there are less guiding principles to derive it. In spite of Yukawa's expectation, the Eq.(\ref{domain}) was not succsseful as an elementary field equation, since it allows several types of ghost solutions. However, if we regard Eq.(\ref{domain}) as an effective-field equation of underlying elementary fields, then the outcome of domain type of field equations is not so extraordinary; and, those field equations are expected to be useful in studying the AdS spacetime itself. The purpose of this work is thus, to investigate the domain type of field equation based on the $\mbox{AdS}_5$ spacetime as an unnoticed route to investigate the AdS${}_5$ spacetime itself and its related problems.

In the next section, we discuss toy models of spin-less particles embedded in five-dimensional Minkowski spacetime, $M_5$, with compact fifth dimensions inclined toward time direction. There, we derive the Green function $G_{ba}$ of the particles embedded in this spacetime; and the propagation of source fields on IR brane is formulated in terms of a reduced Green function $\mathcal{G}_{ba}$ defined out of $G_{ba}$. There, the discussion is also made on the relation between the domain type of fields in IR brane and the fields obtained by the propagation of those source fields.

In section 3, the discussion is extended to the particles embedded in the AdS${}_5$ spacetime with warp factor. In this case, the Green function representing the propagation of particles through the bulk becomes complex. It is, however, shown that the Green function can be reduced essentially to one in the previous section under an appropriate representation of coordinates in addition to a fine tuning of mass parameter.

The domain type of field equations given in those sections depend explicitly on the time direction; and, the Lorentz invariance of the formulation is broken. Section 4 is an attempt to recover the invariance  of the present model by replacing the time direction to a timelike direction associated with a timelike vector on each $y$-fixed brane defined from the spin degrees of freedom of the particles. Finally, section 5 is devoted to the summary and discussion. 

In Appendix A, we discuss the RS spacetime with the extra dimension inclined toward time direction as the solution of Einstein equation.  Appendix B is a short discussion on the spin representation space.

\section{Particles in a five-dimensional spacetime with the extra-dimension inclined to time direction }

We consider a five-dimensional Minkowski spacetime $(M_5)$ $(x^{\hat{\mu}})=(x^\mu,x^5)$,\, $(\mu=0,1,2,3)$ with the metric $\mbox{diag}(\eta_{\hat{\mu}\hat{\nu}})=(-++++)$ realized as the limit of vanishing warp factor in the RS model. Namely, the spacetime is flat, but the fifth dimensional space is the $S^1/Z_2$ orbifold, where $S^1$ is the circle with radius $R$ parametrized by $-\pi\leq \theta\leq \pi$; and we set $x^5=y=\theta R,\,(L=\pi R)$. In other words, the space time is $M_4\otimes S^1/Z_2$. In subsection \ref{2.1}, the Green function of particles in this spacetime is discussed; and, in subsection \ref{2.2}, the discussion is extended to the cases of spacetime with two types of extra dimensions inclined to time direction.

\subsection{Periodic structure of Green function for Klein-Gordon type of equation} \label{2.1}

The Klein-Gordon type of equation for a free spin-less mass $m_0$ particle in this spacetime is
\begin{align}
 \left[ \hat{p}_{\mu} \hat{p}^\mu+\hat{p}_y^2+(m_0c)^2\right]\Psi(x,y)=0\,. \label{KG-function}
\end{align}
Reflecting the $S^1/Z_2$ orbifold structure of the $y$ variable, we require the following Dirichlet type of boundary conditions:
\begin{align}
 \Psi(x,y)\big|_{y=0}=\Psi(x,y)\big|_{y=L}=0. \label{BC}
\end{align}
Then the dynamics of $y$ variable comes to be equivalent to that of a particle in the box $0\leq y\leq L$ (infinite square well); and so, from Eqs.(\ref{KG-function}) and (\ref{BC}), the orthonormal complete basis in the $y$ space are
\begin{align}
\left.
\begin{aligned}
 \phi_n(y) &=\sqrt{\frac{2}{L}}\sin\left(k_ny \right) \\
 \lambda_n &=(\hbar k_n)^2 \label{basis}
\end{aligned}
\right\}\,,~~ 
 \big(\, k_n =\frac{n\pi}{L};\,n=1,2,\cdots \,\big)\,. 
\end{align}
The functions (\ref{basis}) satisfy the eigenvalue equation $\hat{p}^2_y\phi_n(y)=\lambda_n\phi_n(y),\,(\hat{p}_y=-i\hbar\partial_y)$ and the periodic conditions $\phi_n(y)=\phi_n(y+2\pi R)=-\phi_n(-y)$.

The Green function of the Klein-Gordon type of equation (\ref{KG-function}) can be represented by using the complete basis $\{\phi_n\}$ and the eigenstates of four momentum operators defined by $\hat{p}_\mu|p^\prime\rangle=p^{\prime}_\mu |p^\prime\rangle,\,(\, \langle p^\prime|p^{\prime\prime}\rangle=\delta^4(p^\prime-p^{\prime\prime})\,)$,  as
\begin{align}
 G_{ba} &=\langle x_b|\otimes \langle y_b|\left(\hat{p}_{\hat{\mu}}\hat{p}^\mu+\hat{p}^2_y+(m_0c)^2-i\epsilon \right)^{-1}|x_a\rangle\otimes |y_a \rangle \nonumber \\
 &=\frac{i}{2\hbar}\int_0^{\infty}d\tau \langle x_b|\otimes \langle y_b|\left\{\sum_{n=1}^\infty \int d^4p e^{-\frac{i}{\hbar}\tau\frac{1}{2}(p_\mu p^\mu+\lambda_n+(m_0c)^2-i\epsilon)}|p\rangle\langle p|\otimes |\phi_n\rangle \langle\phi_n| \right\} |x_a\rangle\otimes |y_a \rangle \nonumber \\
 &=\frac{i}{2\hbar}\int_0^{\infty}d\tau \sum_{n=1}^\infty \frac{1}{i}\left(\frac{1}{\sqrt{2\pi\hbar i\tau}}\right)^4e^{\frac{i}{\hbar}\frac{1}{2}\left\{\frac{(x_b-x_a)^2}{\tau}-\tau(\lambda_n+(m_0c)^2-i\epsilon)\right\} }\phi_n(y_b)\phi_n(y_a) 
\end{align} 
Here, the summation with respect to $n$ can be rewritten as
\begin{align} 
 \sum_{n=1}^\infty e^{-\frac{i}{\hbar}\frac{\lambda_n}{2}\tau}\phi_n(y_b)\phi_n(y_a) &=\frac{2}{L}\sum_{n=1}^\infty e^{-\frac{i}{\hbar}\frac{(\hbar k_n)^2}{2}\tau}\sin(k_n y_b)\sin(k_n y_a)\nonumber \\
 &=\frac{1}{iL}\sum_{n=-\infty}^\infty e^{-\frac{i}{\hbar}\frac{(\hbar k_n)^2}{2}\tau}\sin(k_n y_b)e^{ik_n y_a} \nonumber \\
 &=\frac{1}{iL}\int dp_y \sum_{n=-\infty}^\infty\delta(p_y-\hbar k_n)e^{-\frac{i}{\hbar}\frac{p_y^2}{2}\tau}\sin\left(\frac{1}{\hbar}p_y y_b\right)e^{\frac{i}{\hbar}p_y y_a}.  \label{n-sum}
\end{align}
Using, further, Poisson's summation rule
\footnote{Since $f(p)\equiv\sum_n\delta\left(p-\frac{n\pi\hbar}{L}\right)$ is a periodic function of $p$ with the period $\pi\hbar/L$, it can be expanded in the Fourier series of $e_r(p)=\sqrt{\frac{L}{\pi\hbar}}e^{i2Lrp/\hbar},\,(r=0,\pm 1,\pm 2 \cdots)$ so that $f(p)=\sum_{r=-\infty}^\infty f_re_r(p)$. Here, the coefficients $\{f_r\}$ are given by $f_r=\int_{-\pi\hbar/L}^{\pi\hbar/L}dpe_r(p)^*f(p)=\sqrt{L/\pi\hbar}$. Substituting the results for the Fourier series, we arrive at the summation rule: $\sum_n\delta\left(p-\frac{n\pi\hbar}{L}\right)=\frac{L}{\pi\hbar}\sum_r e^{i2Lrp/\hbar}$.}
, the right-hand side of Eq.(\ref{n-sum}) becomes
\begin{align}
\mbox{r.h.s} &=\frac{1}{iL}\int dp_y \frac{L}{\pi\hbar} \sum_{r=-\infty}^\infty e^{\frac{i}{\hbar}(2rLp_y)}e^{-\frac{i}{\hbar}\frac{p_y^2}{2}\tau}\sin\left(\frac{1}{\hbar}p_y y_b\right)e^{\frac{i}{\hbar}p_y y_a} \nonumber \\
 &=\frac{-1}{2\pi \hbar}\sum_{r=-\infty}^\infty \int dp_y \left[e^{-\frac{i}{\hbar}\left\{\frac{p_y^2}{2}\tau-(2rL+y_b +y_a)p_y\right\}}-e^{-\frac{i}{\hbar}\left\{\frac{p_y^2}{2}\tau-(2rL-y_b +y_a)p_y\right\}} \right] \nonumber \\
 &=\sum_{r=-\infty}^\infty \frac{-1}{\sqrt{2\pi\hbar i\tau}}\left[e^{\frac{i}{\hbar}\frac{(2rL+2\breve{y}_{ba})^2}{2\tau}}-e^{\frac{i}{\hbar}\frac{(2rL-\bar{y}_{ba})^2}{2\tau}} \right]\,, 
\end{align}
where the use has been made of the notations
\begin{align}
 \bar{f}_{ba} &=f_b-f_a  & \cdots &~\mbox{relative coordinate} \\
 \breve{f}_{ba} &=\frac{1}{2}(f_b+f_a)  & \cdots &~\mbox{intermediate coordinate}
\end{align}
Therefore, the Green function $G_{ba}$ can be represented by \cite{Box1,Box2}
\begin{align}
 G_{ba} &=\frac{i}{2\hbar}\int_0^{\infty}d\tau\sum_{r=-\infty}^\infty\frac{-1}{i}\left(\frac{1}{\sqrt{2\pi\hbar i\tau}}\right)^5 e^{-\frac{i}{2\hbar}\{(m_0c)^2-i\epsilon\}\tau} \times \nonumber \\
 & \times \left[e^{\frac{i}{2\hbar}\frac{1}{\tau}\left\{(\bar{x}_{ba})^2+(2rL+2\breve{y}_{ba})^2\right\}}- e^{\frac{i}{2\hbar}\frac{1}{\tau}\left\{(\bar{x}_{ba})^2+(2rL-\bar{y}_{ba})^2\right\}} \right]  \nonumber \\
 &=\sum_{r=-\infty}^\infty\left[ K\left((\bar{x}_{ba})^2+(2rL+2\breve{y}_{ba})^2 \right)-K\left((\bar{x}_{ba})^2+(2rL-\bar{y}_{ba})^2\right)\right]\,, \label{Green-1}
\end{align}
where
\footnote{
If we use $\int_0^\infty \frac{d\tau}{\tau^{5/2}}e^{i\left(\frac{a}{\tau}-b\tau\right)}=-\sqrt{\frac{\pi}{ib}}\frac{1}{a^2}\left(\frac{1}{2}\sqrt{ab}+ab\right)e^{-2\sqrt{ab}},(a,b>0)$, one can write $K(z)=i\frac{(m_0c)^3}{\pi^2(2\hbar)^5}f\left[z\left(\frac{m_0c}{2\hbar}\right)^2\right]$, where $f(x)=\left(\frac{1}{2}\sqrt{x}+x\right)e^{-2\sqrt{x}}/x^2$.
}

\begin{align}
 K(z) &=\frac{-1}{2\hbar}\int_0^{\infty}d\tau \left(\frac{1}{\sqrt{2\pi\hbar i\tau}}\right)^5e^{\frac{i}{2\hbar}\left[\frac{z}{\tau}-\tau((m_0c)^2)\right]}. \label{Green-2}
\end{align}
It is obvious that the Green function $G_{ba}(\bar{x}_{ba},\bar{y}_{ba},\breve{y}_{ba})$ satisfies the boundary conditions $G_{ba}\big|_{y_a=0,L}=G_{ba}\big|_{y_b=0,L}=0$ as a result of Eq.(\ref{BC}). We also note that the $G_{ba}(\bar{x}_{ba},\bar{y}_{ba},\breve{y}_{ba})$ is not invariant under the translation to the fifth direction due to the presence of $\breve{y}_{ba}$, the mean value of $y_a$ and $y_b$, but invariant under the discrete transformations $\breve{y}_{ba}\rightarrow \breve{y}_{ba}\pm L$ and $\bar{y}_{ba}\rightarrow \bar{y}_{ba}\pm 2L$; and so, one can write
\begin{align}
 G_{ba} &(\bar{x}_{ba},\bar{y}_{ba},\breve{y}_{ba}) =G_{ba}(\bar{x}_{ba},\bar{y}_{ba},\breve{y}_{ba}\pm L)=G_{ba}(\bar{x}_{ba},\bar{y}_{ba}\pm 2L,\breve{y}_{ba}). \label{periodicity}
\end{align}
Then, the field defined by
\begin{align}
 \Psi(x_b,y_b)=\int d^4x_a\int dy_a G_{ba}(\bar{x}_{ba},\bar{y}_{ba},\breve{y}_{ba})\Phi(x_a,y_a) \label{a-to-b}
\end{align}
satisfies the Klein-Gordon type of equation with the initial source field $\Phi(x_a,y_a)$; that is, $(\hat{p}_\mu\hat{p}^\mu+\hat{p}_y^2+(m_0c)^2)\Psi=\Phi$. By virtue of Eq.(\ref{periodicity}), the field $\Psi(x_b,y_b)$ also satisfies the periodic condition $\Psi(x_b,y_b\pm 2L)=\Psi(x_b,y_b)$. 

It should be stressed that the $\Psi(x_b,y_b)$ is a one-particle wave function, a classical field, resulting from the propagation of $\Phi(x_a,y_a)$ under the effect of the fifth dimension; in other words, $\Psi(x_b,y_b)$ is a functional of $\Phi(x_a,y_a)$ with a dressed effect in the fifth dimension. Then, the field $\Psi(x_b,y_0),\,(0<y_0<L)$ with a source field $\Phi(x_a,y_a)=\delta(y_a-y_0)\Phi_a(x_a)$ is a dressed functional of $\Phi_a(x_a)$ on a fixed $y\,(=y_0)$ plane. 

It is, now, interesting to read $y_0=0$ and $y_0=L$ planes as UV and IR branes respectively for realistic applications, despite the warp factor is not introduced in this stage. Here, if we can put $y_0=L$, then $\Phi_a(x_a)$ and $\Psi_b[\Phi_a,x_b]=\Psi_b(x_b,L)$ will be respectively a source field and its dressed functional living in IR brane. This viewpoint looks to cause a problem, since the Green function vanishes on $y_0=0,L$ planes. When the extra dimension is inclined toward time direction, however, the situation will be changed. In such a case, as will be shown in the next subsection, we don't have to identify the ends of $y$ space and the points, on which the boundary conditions (\ref{BC}) are applied. In addition to this, then, the Green function has a timelike periodicity through the $y$ periodicity, which add to $\Psi_b[\Phi_a,x_b]$ a similar aspect to Yukawa's domain type of field. To confirm this conjecture, in what follows, we discuss two types of extra dimensions inclined toward time direction. \vspace{1mm}\\

\subsection{Models of the fifth dimension inclined toward time direction} \label{2.2}

In the following attempts (i) and (ii), each extra-dimension is inclined toward time direction through linear-coordinate transformations with constant coefficients. Although the introduction of such an extra dimension spoils the Lorentz symmetry of the formalism, we don't worry about this problem until section 4, where a model recovering the symmetry is discussed. \\

\noindent
{(i)~\bf Lorentz boost} \vspace{2mm}\\
Let us consider two reference systems in five-dimensional spacetime: $\Sigma$ system with the coordinates $(x^{\hat{\mu}})=(x^{0},x^{i},x^5),\,(x^5=y)$ and $\Sigma^\prime$ system with the coordinates $(x^{\prime\hat{\mu}})=(x^{\prime 0},x^{\prime i},x^{\prime 5}),\,(x^{\prime 5}=y^\prime)$, which moves against the $\Sigma$ system along the fifth direction with a constant velocity $c\beta$; we set the coincidence $\Sigma=\Sigma^\prime$ at $x^0=0$. Using the basis vector $e^{\hat{\mu}}$ toward $\hat{\mu}$-th direction with the covariant components $(e^{\hat{\mu}})_{\hat{\nu}}=\delta^{\hat{\mu}}_{\hat{\nu}}$, one can write $x^{\hat{\mu}}=e^{\hat{\mu}}\cdot x=(e^{\hat{\mu}})_{\hat{\rho}}x^{\hat{\rho}}$. Similarly, in terms of the basis vector defined by
\begin{align}
\left.
\begin{aligned}
 e^{\prime\, 0} &=C_\beta e^{0}-S_\beta e^5 \\
 e^{\prime\, i} &=e^{i} \\
 e^{\prime\, 5} &=C_\beta e^5-S_\beta e^0 \\
\end{aligned}\right\} 
~~\left(\,C_\beta=1/\sqrt{1-\beta^2}, S_\beta=\beta/\sqrt{1-\beta^2} \,\right), \label{boost}
\end{align}
the coordinates in $\Sigma^\prime$ system can be written as $x^{\prime\hat{\mu}}=e^{\prime\,\hat{\mu}}\cdot x$. By definition, the $\{e^{\prime\, \hat{\mu}} \}$ is a moving frame against the static frame $\{e^{\hat{\mu}}\}$.

Now, writing the particle position variables as $x^{\hat{\mu}}(\tau)$ with the time ordering parameter $\tau$, the square of the world-line interval of a particle becomes
\begin{align}
 ds^2=\eta_{\hat{\mu}\hat{\nu}}\delta x^{\hat{\mu}}\delta x^{\hat{\mu}}~~(\, \delta x^{\hat{\mu}}=\dot{x}^{\prime\hat{\mu}}d\tau=(e^{\prime\hat{\mu}}\cdot\dot{x})d\tau \,), \label{ds-i}
\end{align}
where $\dot{x}^{\hat{\mu}}\,(\dot{x}^{\prime\hat{\mu}})$ is the five velocity of the particle in $\Sigma\,(\Sigma^\prime)$ system. Then, the Lagrangian of a mass $m_0$ particle is given by
\begin{align}
 \mathcal{L}=\frac{1}{2}\left\{\frac{1}{e}\left(\frac{ds}{d\tau}\right)^2-(m_0c)^2e\right\}. \label{Lagrangian}
\end{align}
Here the $e$ is the einbein in $\tau$ space. The variation of $\mathcal{L}$ with respect to $e$ leads to the constraint
\begin{align}
 \mathcal{K} &=\eta^{\hat{\mu}\hat{\nu}}p^\prime_{\hat{\mu}}p^\prime_{\hat{\nu}}+(m_0c)^2=\eta^{\hat{\mu}\hat{\nu}}p_{\hat{\mu}}p_{\hat{\nu}}+(m_0c)^2=0
\end{align}
where $p_{\hat{\mu}}\,(p^\prime_{\hat{\mu}})$ is the momentum conjugate to $x^{\hat{\mu}}\,(x^{\prime{\hat{\mu}}})$.

\begin{figure}
\begin{minipage}{6cm}
 \includegraphics[width=6cm]{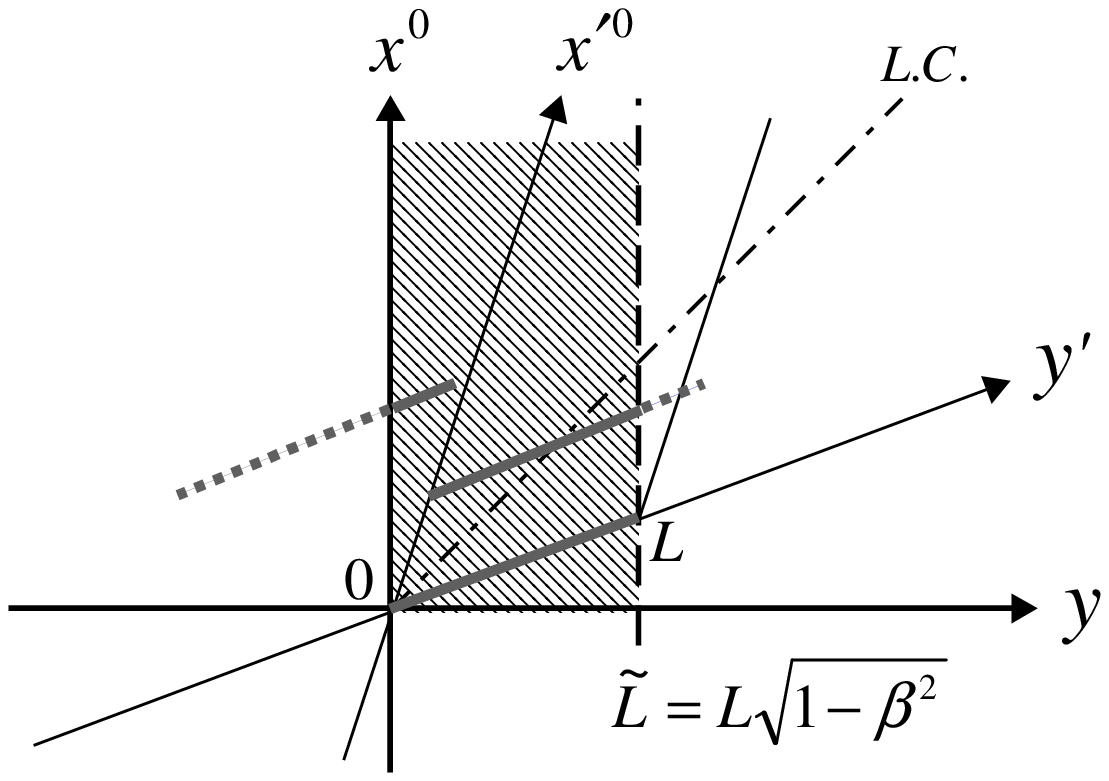}
\end{minipage}
\begin{minipage}{6cm}
Since the fifth dimension is compact, the Lorentz boost should be understood in a covering space of $0\leq y^\prime \leq L$ $(0\leq y \leq \tilde{L})$ extended by the identification $y^\prime\equiv y^\prime+L$ $(y \equiv y+\tilde{L})$. Then, the time dependent $y^\prime$ axis is represented equivalently by moving oblique lines, like the patterns of barber's pole, in the $(x^0,y)$ plane with $\,0\leq y\leq \tilde{L}$. 
\end{minipage}
\caption{Lorentz boost to the direction of $y$ axis}
\label{Lorentz-boost}
\end{figure}

In q-number theory, the wave function characterized by $\hat{\mathcal{K}}|\Psi\rangle=0$ is solved as a function of $x^{\hat{\mu}}\,(x^{\prime\hat{\mu}})$ in the reference system $\Sigma\,(\Sigma^\prime)$. 
The stand points of two reference systems are relative; and we regard Eq.(\ref{KG-function}) as one in $\Sigma^\prime$ system. The extra-dimension in $\Sigma^\prime$ system has the interval $0\leq y^\prime\leq L$, in which the boundary conditions (\ref{BC}) is. On the other side, the interval of the extra-dimension is $0\leq y\leq\tilde{L}=L\sqrt{1-\beta^2}$ for the observer in $\Sigma$ system (Fig.\ref{Lorentz-boost}).

Under those understandings, the Green function $G_{ba}(\bar{x}^\prime_{ba},\bar{y}^\prime_{ba},\breve{y}^\prime_{ba})$ in $\Sigma^\prime$ system is the function (\ref{Green-1}) written by $x^{\prime\hat{\mu}}$. The Green function in $\Sigma$ system is, then,  obtained from that in $\Sigma^\prime$ system by substituting $e^{\prime\,\hat{\mu}}\cdot x$ for $x^{\prime\,\hat{\mu}}$ so that
\begin{align}
 G_{ba}=G_{ba}(\bar{x}^0_{ba}C_\beta-\bar{y}_{ba}S_\beta,  \bar{x}^i_{ba},\bar{y}_{ba}C_\beta-\bar{x}^0_{ba}S_\beta,\breve{y}_{ba}C_\beta-\breve{x}^0_{ba}S_\beta). \label{Green-Sigma}
\end{align}
The resultant expression says that the $G_{ba}$ is a periodic function of $y$ with the period $\tilde{L}$; and, in this sense, we can put the congruence $y\equiv y+\tilde{L}$ in $y$ space. Under this congruence, as shown in Fig.\ref{Lorentz-boost}, we can regard $y=0$ and $y=\tilde{L}$ as the boundaries of extra dimension in $\Sigma$ system, which should be identified as UR and IR ends respectively.

The propagation of initial source field $\Phi(x_a,y_a)=\delta(y_a-\tilde{L})\Phi_a(x_a)$ on $y=\tilde{L}$ plane is, thus, described by the reduced Green function
\begin{align}
 \mathcal{G}^{(B)}_{ba}=G_{ba}\big|_{y_a=y_b=\tilde{L}}=G_{ba}(\bar{x}^0_{ba}C_\beta,  \bar{x}^i_{ba},-\bar{x}^0_{ba}S_\beta,\tilde{L}C_\beta-\breve{x}^0_{ba}S_\beta),
\end{align}
by which one can write the counterpart of Eq.(\ref{a-to-b}) as
\begin{align}
 \Psi_b[\Phi_a,x_b]=\Psi(x_b,\tilde{L})=\int d^4x_a\mathcal{G}^{(B)}_{ba}\Phi_a(x_a). \label{a-to-b-boost}
\end{align}
This equation defines a mapping between the fields $\Phi_a$ and $\Psi_b$ in IR brane within the framework of c-number theory. The notation $\Psi_b[\Phi_a,x_b]$ says that $\Psi_b$ is a functional of $\Phi_a$ simultaneously with a function of $x_b$; we can view that $\Psi_b$ is a dressed field of $\Phi_a$ in the effect of the extra dimension. When $\Phi_a$ is a quantized field, then $\Psi_b$ will be a q-number field as a functional of $\Phi_a$. 

Now, by taking $y\equiv y+\tilde{L}$ into account, Eq.(\ref{a-to-b-boost}) with $L_B=\tilde{L}/S_\beta$ leads to
\begin{align}
\hspace{-5mm}
 e^{\pm L_B(\partial_0)_b}\Psi_b[\Phi_a,x_b] &=\int d^4x_a\left\{e^{\pm L_B(\partial_0)_b}\mathcal{G}^{(B)}_{ba}\right\}\Phi_a(x_a)  \nonumber \\
 &=\int d^4x_a\left\{e^{\mp L_B(\partial_0)_a} G_{ba}(\bar{x}_{ba}^0C_\beta,\bar{x}_{ba}^i,-\bar{x}^0_{ba}S_\beta,\tilde{L}C_\beta-\breve{x}^0_{ba}S_\beta \mp \tilde{L}) \right\}\Phi_a(x_a) \nonumber \\
 &=\int d^4x_a \mathcal{G}^{(B)}_{ba}\left\{e^{\pm L_B(\partial_0)_a}\Phi_a(x_a)\right\} \nonumber \\
 &=\int d^4x_a\mathcal{G}^{(B)}_{ba}\Phi_a(x^0_a\pm L_B, x^i_a). \label{displacement-LB}
\vspace{3mm}
\end{align}
The result is not trivial, since this finite-time displacement equation holds only for the parameter $L_B=L/S_\beta$ with $\beta\neq 0$; otherwise, the form of the Green function $\mathcal{G}^{(B)}_{ba}$ in Eq.(\ref{displacement-LB}) will be changed.  \\

\noindent
{\bf (ii)~ rotating $S^1$ circle}\vspace{2mm}

As another possible model of the extra dimension inclined toward a time direction, let us consider the circle $S^1$ with radius $R$, on which two coordinate systems of angle variables $\{\theta \}$ and $\{\theta^\prime\}$ are defined. First, the $\{\theta\}$ defines the fifth-dimensional space with the period $2\pi$ by $\theta\equiv \theta+2\pi$. Secondly, the $\{\theta^\prime\}$ is the angle variable defined by $\theta=\theta^\prime+\Omega x^0$, where $c\Omega$ is a constant angular velocity of the rotation. 
\begin{figure}
\centering
 \includegraphics[width=7cm]{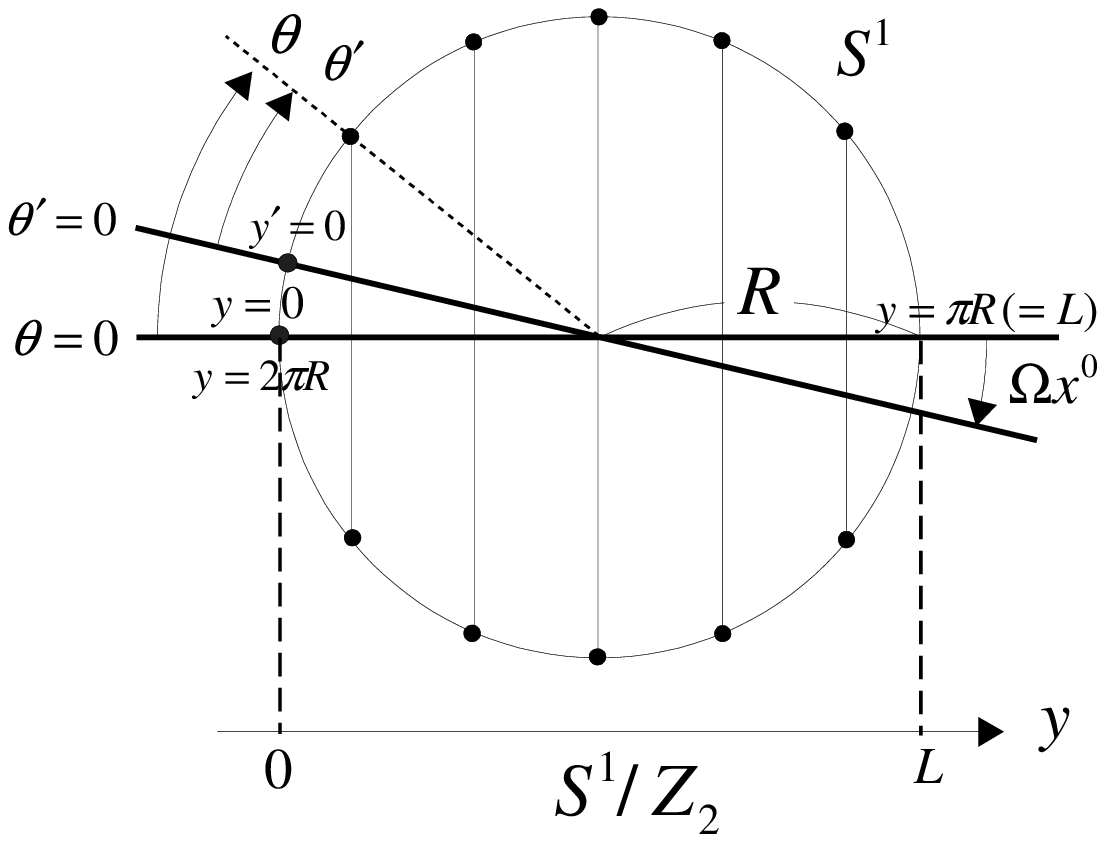}
\caption{Coordinates $\theta$ and $\theta^\prime$ on a ring}
\label{ring}
\end{figure}

Under the projection from the circle to horizontal axis by $dy=Rd\theta$ and $dy^\prime=Rd\theta^\prime(=dy-\beta dx^0,\,\beta=\Omega R)$ as in Fig.2, one can regard the coordinates $(x^\mu,y)$ and $(x^\mu,y^\prime)$ respectively correspond to those in $\Sigma$ and $\Sigma^\prime$ systems in the case (i), although those are not inertial reference frames. In this case, the basis vector of $\Sigma^\prime$ system is $(e^{\prime\,\mu},e^{\prime\,5})=(e^{\mu},e^5-\beta e^{0})$ with $(e^{\hat{\mu}})_{\hat{\nu}}=\delta^{\hat{\mu}}_{\hat{\nu}}$.
 Then, as in Eq.(\ref{ds-i}), the world-line interval of a particle can be set as
\begin{align}
 ds^2 &=\eta_{\hat{\mu}\hat{\nu}}\delta x^{\hat{\mu}}\delta x^{\hat{\nu}}~~(\delta x^{\hat{\mu}}=(e^{\prime\,\hat{\mu}}\cdot\dot{x})d\tau). \label{ds-ii}
\end{align}
Since, by definition $\theta\equiv \theta+2\pi$ and $\theta^\prime\equiv \theta^\prime+2\pi$, one can set $y\equiv y+2L$ and $y^\prime\equiv y^\prime+2L$, where $L=\pi R$. 
If we require $y\equiv y+L\equiv -y$ for fields on $y$ space, the circle becomes the orbifold $S^1/Z_2$. 

The Lagrangian of a mass $m_0$ particle embedded in this spacetime has the same form as (\ref{Lagrangian}) with the world-line interval (\ref{ds-ii}). The variation of $\mathcal{L}$ with respect to $e$ yields the constraint
\begin{align}
 \mathcal{K}\equiv p^{\prime\hat{\mu}}p^\prime_{\hat{\mu}}+(m_0c)^2=-(p_0+\beta p_y)^2+p^ip_i+p_y^2+(m_0c)^2=0 \,,
\end{align}
where $p^\prime_{\hat{\mu}}\,(p_{\hat{\mu}})$ is the momentum conjugate to $x^{\prime\hat{\mu}}\,(x^{\hat{\mu}})$. In q-number theory, $\hat{\mathcal{K}}\Psi=0$ represents the field equation of this particle. As in the case of (i), the Green function $G_{ba}=(\hat{\mathcal{K}}^{-1})_{ba}$ associated with the boundary conditions $G_{ba}\big|_{y^\prime=0,L}=0$ is given by Eq.(\ref{Green-1}) in $x^{\prime\hat{\mu}}$ coordinates. Substituting $(x^\mu,y-\beta x^0)$ for $(x^{\prime\mu},y^\prime)$, the Green function in $x^{\hat{\mu}}$ coordinates, thus, can be written as
\begin{align}
 G_{ba}=G_{ba}(\bar{x}_{ba},\bar{y}_{ba}-\beta\bar{x}^0_{ba},\breve{y}_{ba}-\beta\breve{x}^0_{ba}).
\end{align}
Here, we again regard that the hyper-surfaces of $y=0$ and $y=L$ are corresponding to the UV and IR branes, respectively. The propagation of initial source field $\Phi(x_a,y_a)=\delta(y_a-L)\Phi_a(x_a)$ on $y=L$ brane is described by the reduced Green function
\begin{align}
 \mathcal{G}^{(R)}_{ba}=G_{ba}\big|_{y_a=y_b=L}=G_{ba}\left(\bar{x}^\mu_{ba},-\beta\bar{x}^0_{ba},L-\beta \breve{x}^0_{ba}\right),
\end{align}
so that
\begin{align}
 \Psi_b[\Phi_a,x_b]=\Psi(x_b,L)=\int d^4x_a\mathcal{G}^{(R)}_{ba}\Phi_a(x_a). \label{a-to-b-rotation}
\end{align}
Then, in a similar way discussed in Eq.(\ref{displacement-LB}), one can derive the finite-time displacement equation
\begin{align}
 e^{\pm L_R(\partial_0)_b}\Psi_b[\Phi_a,x_b] &=\int d^4x_a\mathcal{G}^{(R)}_{ba}\Phi_a(x^0_a\pm L_R, x^i_a), \label{displacement-R}
\end{align}
where $L_R=L/\beta$. The result is the same as Eq.(\ref{displacement-LB}) except the substitution $(\mathcal{G}^{(B)}_{ba},L_B)\rightarrow (\mathcal{G}^{(R)}_{ba},L_R)$. So, as the occasion arises to lump the cases (i) and (ii), we write $L_B$ and $L_R$ as $L_e$ in common; and we use the symbol $\mathcal{G}_{ba}$ simply.

\subsection{Free domain type of field equation in IR brane}

Until now, we have only shown that some types of extra dimensions inclined toward time direction lead to the finite-time displacement equations such as
\begin{align}
  e^{\pm L_e(\partial_0)_b}\Psi_b[\Phi_a,x_b]=\int d^4x_a\mathcal{G}_{ba}\Phi_a(x^0_a\pm L_e, x^i_a). \label{displacement}
\end{align}
In order to relate this equation to Yukawa's domain type of field equation, let us consider the case that $\Psi_b$ satisfies the closed equation on IR brane
\begin{align}
 e^{L_e(\partial_0)_b}\Psi_b[\Phi_a,x_b]=e^{-\frac{i}{\hbar}S(\hat{\bm{p}}_b)}\Psi_b[\Phi_a,x_b], \label{domain-1}
\end{align}
as a simple model of Eq.(\ref{domain}). Since $\mathcal{G}_{ba}$ is a function of $(\bar{x}^{\hat{\mu}},\breve{x}^0)$, Eq.(\ref{displacement}), Eq.(\ref{domain-1}), and the integration by parts with respect to $x^i_a$ give rise to
\begin{align}
 \Phi_a(x^0_a+ L_e, x^i_a)=e^{-\frac{i}{\hbar}S(\hat{\bm{p}}_a)}\Phi_a(x_a), \label{suitable source}
\end{align}
or
\begin{align}
  e^{-\frac{i}{\hbar}L_e\left(\hat{p}^0-\frac{1}{L_e}S(\hat{\bm{p}})\right)}\Phi_a(x)=\Phi_a(x). \label{domain-2}
\end{align}
The Eq.(\ref{domain-2}) can be satisfied by
\begin{align} 
 \Phi_a(x)=\theta(\hat{p}^0)\delta\left(\hat{p}^{02}-\frac{1}{L_e^2}S(\hat{\bm{p}})^2 \right)\phi_a(x), \label{domain-S}
\end{align}
where $\phi_a(x)$ is an arbitrary function of $x^\mu$. The function of $\hat{p}^\mu$ in front of $\phi_a(x)$ in Eq.(\ref{domain-S}) is not Lorentz invariant in general; however, for $\frac{1}{L_e}S(\hat{\bm{p}})=\sqrt{\hat{\bm{p}}^2+(m_0c)^2}$, Eq.(\ref{domain-S}) becomes the positive frequency part of a free on-mass-shell field such that  
\begin{align}
 \Phi_a(x)=\theta(\hat{p}^0)\delta(\hat{p}^2+(m_0c)^2)\phi_a(x). \label{source}
\end{align}
Then, back to Eq.(\ref{domain}), the field $\Psi_b$ with the initial source (\ref{source}) satisfies the following domain type of field equation
\begin{align}
 e^{L_e(\partial_0)_b}\Psi_b[\Phi_a,x_b] =e^{-\frac{i}{\hbar}L_e\sqrt{\hat{\bm{p}}_b^2+(mc)^2}}\Psi_b[\Phi_a,x_b].  \label{domain-3}
\end{align}
If we choose, here, the negative frequency part of free on-mass-shell field $\Phi_a$ by substituting $\theta(\hat{p}^0)\rightarrow \theta(-\hat{p}^0)$ in Eq.(\ref{source}), then Eq.(\ref{domain-2}) and Eq.(\ref{domain-1}) will be realized obviously for $\frac{1}{L_e}S(\hat{\bm{p}})=-\sqrt{\hat{\bm{p}}^2+(m_0c)^2}$. Since the $\Psi_b$ is a functional of free field $\Phi_a$, Eq.(\ref{domain-3}) should be read as a free domain type of field equation.

It must be noted that the initial source $\Phi_a$ is arbitrary in the stage of the finite-time displacement equation (\ref{displacement}). In order to make Eq.(\ref{displacement}) to a closed form of $\Psi_b$, some conditions on $\Phi_a$ such as (\ref{source}) come to be necessary. Once Eq.(\ref{domain-1}) is satisfied by $\Psi_b$, then the functional $\Psi_c[\Psi_b,x_c]$ satisfies the domain type of field equation
\begin{align}
 e^{L_e(\partial_0)_c}\Psi_c[\Psi_b,x_c] =e^{-\frac{i}{\hbar}S(\hat{\bm{p}}_c)}\Psi_c[\Psi_b,x_c]. \label{domain-4}
\end{align}
Since Eq.(\ref{domain-4}) is a closed equation for $\Psi_b$'s, the functional space of $\Psi_b$'s can be defined as a solution space of Eq.(\ref{domain-4}). 

Further, in general, the functional relation between $\Psi_b$ and $\Phi_a$ spoils the symmetry inherent in $\Phi_a$
\footnote{ In $m_0=0$ limit, $\Phi_a(x)$ may have a definite scale dimension provided $\phi_a(x)$ has a definite scale dimension. Namely, if $\phi_a(\lambda x)=\lambda^{-d}\phi_a(x)$, then $\Phi_a(\lambda x)=\lambda^{-(d-2)}\Phi_a(x)$.  Even in this case, $\Psi_b$ does not have a definite scale dimension due to the presence of $L_e$ in $G_{ba}$.
 }
; for example, Eq.(\ref{domain-3}) is not Lorentz invariant even if the $\Phi_a$ in Eq.(\ref{source}) is a scalar function. In our line of approach, the recovery of the Lorentz invariance is a technical problem, which will be done in section 4.

\section{Spin-less particles in AdS${}_5$ spacetime}

In this section, we discuss a free spin-less particle embedded in a modified $\mbox{AdS}_5$ spacetime characterized by the warp factor $e^{-2ky}$.  In the RS model, the spacetime  is characterized by the metric
\begin{align}
 \mbox{diag}(g_{\hat{\mu}\hat{\nu}})=(e^{-2ky}\eta_{\mu\nu},1)  ~~(\, x^5=y \,). \label{warp}
\end{align}
Then, as usual, the square of the world-line interval of a particle embedded in this spacetime is given by $ds^2=g_{\hat{\mu}\hat{\nu}}\delta x^{\hat{\mu}}\delta x^{\hat{\nu}}$, where $\delta x^{\hat{\mu}}=(e^{\hat{\mu}}\cdot\dot{x})d\tau$ with $(e^{\hat{\mu}})_{\hat{\nu}}=\delta^{\hat{\mu}}_{\hat{\nu}}$. If we, here, replace the $(e^{\hat{\mu}}\cdot\dot{x})$ by $(e^{\prime\,\hat{\mu}}\cdot\dot{x})$ with $e^{\prime\, \hat{\mu}}=(e^\mu,e^5-\beta e^0)$, the case (ii) basis vector in subsection \ref{2.2}, then the metric $g^\prime_{\hat{\mu}\hat{\nu}}=g_{\hat{\alpha}\hat{\beta}}(e^{\prime\hat{\alpha}})_{\hat{\mu}}(e^{\prime\hat{\beta}})_{\hat{\nu}}$ will take on an effect of the rotating extra-dimension. One can regard alternatively the $(e^{\prime\,\hat{\mu}}\cdot\dot{x})$ as an effective velocity of the particle relative to the background  spacetime characterized by $g_{\hat{\mu}\hat{\nu}}$ as in the rigid body motion. It is of course that the wave equation obtained from the Lagrangian (\ref{Lagrangian}) does not depend on those standpoints. 

In the following, we deal with a modified world-line interval such as
\footnote{In order to know the background spacetime of the particle in more detail, it is interesting to study spacetime with the metric $g^\prime_{\hat{\mu}\hat{\nu}}=g_{\hat{\alpha}\hat{\beta}}(e^{\prime\,\hat{\alpha}})_{\hat{\mu}}(e^{\prime\,\hat{\beta}})_{\hat{\nu}}$ as a solution of the Einstein equation. From this purpose, the discussion will be made on the spacetime characterized by the line element
\[ ds^2=e^{-2ky}\{-(Adx^0+Bdy)^2+dx^idx^i\}+(Cdy+D(y)dx^0)^2 \]
in appendix A.}
\begin{align}
\begin{split}
 ds^2 &=g_{\hat{\mu}\hat{\nu}}\delta x^{\hat{\mu}}\delta x^{\hat{\nu}}~~(\, \delta x^{\hat{\mu}}=(e^{\prime\hat{\mu}}\cdot\dot{x})d\tau \,), \\
 (e^{\prime\,\hat{\mu}}) &=(e^{\mu}, e^5 -\beta e^{-ky}e^0) \label{beta-frame}
\end{split}
\end{align}
to derive a well behaved wave equation in terms of the variable $z=e^{-ky}$. The introduction of a five vector such as (\ref{beta-frame}) spoils the Lorentz invariance in each $y$-fixed  brane; however, we don't worry about this problem, since the invariance is easily recovered by considering a particle with dynamical variables other than the position variables as shown in the next section.
  
\subsection{Wave equation of the particle and the tuning of its mass in UV brane} \label{Tuning}

We start with the Lagrangian (\ref{Lagrangian}) with the world-line interval (\ref{beta-frame}). As usual, the variation of this Lagrangian with respect to $e$ gives rise to the constraint
\begin{align}
 e^{2ky}\left\{-(p_0+\beta e^{-ky}p_y)^2+p^ip^i\right\}+\left\{p_y^2+(m_0c)^2\right\}=0. \label{starting equation}
\end{align}
From a technical reason for transition to q-number theory, we divide this constraint into the following two constraints in extended phase space:
\begin{align}
 \mathcal{K}_\beta &\equiv e^{2ky}\left\{-(p_0-p_u)^2+p^ip^i\right\}+\left\{p_y^2+(m_0c)^2\right\}=0, \label{K-beta} \\
 \phi_\beta &\equiv p_u +\beta e^{-ky}p_y=0, \label{u-constraint}
\end{align}
where $u$ is an auxiliary variable in the extended phase space, and $p_u$ is the momentum conjugate to $u$
\footnote{If we introduce the auxiliary variable $u$ in the beginning, the Lagrangian (\ref{Lagrangian}) with 
\[ ds^2=e^{-2ky}\eta_{\mu\nu}dx^\mu dx^\nu+\left\{dy-\beta e^{-ky}(dx^0+du) \right\}^2 \]
is able to derive directly the constraints (\ref{K-beta}) and (\ref{u-constraint}).}
. The Eqs.(\ref{K-beta}) and (\ref{u-constraint}) are second class constraints; and so, if we eliminate $p_u$ by regarding Eq.(\ref{u-constraint}) as its definition, then Eq.(\ref{K-beta}) will be reduced to the constraint (\ref{starting equation}).

Now, in q-number theory, $\hat{\mathcal{K}}_\beta\Psi=0$ and $\hat{\phi}_\beta\Psi=0$ represent the master wave equation and its supplementary condition. Writing $\hat{\mathcal{K}}_{(\beta)}=e^{-2ky}\hat{\mathcal{K}}_\beta$, we use those equations in the following forms:
\begin{align}
 \hat{\mathcal{K}}_{(\beta)} \Psi &= \left[ \left\{-(\hat{p}_0-\hat{p}_u)^2+\hat{p}^i\hat{p}^i\right\}+e^{-2ky}\left\{\hat{p}_y^2+(m_0c)^2\right\}\right]\Psi=0, \label{K-beta-2} \\
 \hat{\phi}_\beta\Psi &=\left(\hat{p}_u +\beta e^{-ky}\hat{p}_y\right)\Psi=0. \label{u-constraint-2}
\end{align}
The Klein-Gordon type of equation (\ref{K-beta-2}) says that $m_y=m_0e^{-ky}$ is the effective mass of the particle in a $y$ fixed brane; and so, $m_0$ and $m_L$ are masses in UV and IR branes respectively. To deal with Eq.(\ref{K-beta-2}), it is convenient to use the variable $z=e^{ky}\,,(1\leq z\leq L_z=e^{kL})$ rather than $y$. Then, by taking
\begin{align}
 \hat{p}_y^2=z^2\left\{\left(\hat{p}_z-i\frac{\hbar k}{2z}\right)^2-\frac{(\hbar k)^2}{4z^2}\right\},~\left(\,\hat{p}_z=-i(\hbar k)\frac{\partial}{\partial z}\,\right),  \label{py-square}
\end{align}
into account, Eqs.(\ref{K-beta-2}) and (\ref{u-constraint-2}) can be rewritten as follows
\begin{align}
 \hat{\mathcal{K}}_{(\beta)}\Psi &=\left[\left\{-(\hat{p}_0-\hat{p}_u)^2+\hat{p}_i\hat{p}^i\right\}+\left(\hat{p}_z-i\frac{\hbar k}{2z}\right)^2-(\hbar k)^2\frac{\Delta}{z^2}\right]\Psi=0,  \label{K-beta-3} \\
 \hat{\phi}_\beta\Psi &= \left(\hat{p}_u +\beta \hat{p}_z\right)\Psi=0, \label{u-constraint-3}
\end{align}
where
\begin{align}
 \Delta =\left(\frac{1}{2}\right)^2-\left(\frac{m_0 c}{\hbar k}\right)^2\,.
\end{align}
Hereby, under the unitary transformations $\hat{\mathcal{K}}^{\prime}_{(\beta)}=U\hat{\mathcal{K}}_{(\beta)}U^{-1}$ and $\hat{\phi}_\beta^\prime=U\hat{\phi}_\beta U^{-1}$ with $U=e^{-\frac{i}{\hbar}x^0\hat{p}_u}$,  those equations become
\begin{align}
\hat{\mathcal{K}}^\prime_{(\beta)}\Psi^\prime &=\left[\hat{p}_\mu\hat{p}^\mu+\left(\hat{p}_z-i\frac{\hbar k}{2z}\right)^2-(\hbar k)^2\frac{\Delta}{z^2}\right]\Psi^\prime=0, \label{K-prime}\\
 \hat{\phi}_\beta^\prime\Psi^\prime &=\left( \hat{p}_u +\beta \hat{p}_z \right)\Psi^\prime=0, \label{u-prime}
\end{align}
where $\Psi^\prime=U\Psi$. In this stage, the $\hat{p}_u$ no longer remains in $\hat{\mathcal{K}}^\prime_{(\beta)}$; and so, the $\hat{p}_u$ can be removed out of formalism by regarding (\ref{u-prime}) as the definition $\hat{p}_u\equiv -\beta\hat{p}_z$. In relation to this, however, we have to read $U=e^{\frac{i}{\hbar}x^0\beta\hat{p}_z}$.
Thus, hereafter, we may deal with Eq.(\ref{K-prime}) as the wave equation for the particle under consideration; furthermore, if we carry out the transformation $\tilde{\Psi}=\sqrt{z}\Psi^\prime$, the Eq.(\ref{K-prime}) becomes a simpler form
\begin{align}
 \hat{\mathcal{K}}_{[\beta]}\tilde{\Psi}=\left[\hat{p}^\mu\hat{p}_\mu+\hat{p}_z^2-(\hbar k)^2\frac{\Delta}{z^2}\right]\tilde{\Psi}=0~~(1\leq z\leq z_L)\,. \label{master-wave}
\end{align}

Now, Eq.(\ref{master-wave}) has an interesting form, in which $V(z)\equiv -(\hbar k)^2\frac{\Delta}{z^2}$ plays the role of an effective potential in $z$-space, the sign of which is decided by the ratio between $m_0$ and $k$. If we assume that the large mass hierarchy is realized by $M_W\simeq e^{-kL}M_{P}$, where $M_W$ and $M_P$ are respectively a mass in TeV scale and the Planck mass, then the spacetime parameter will be estimated as $kL\simeq 35$. The $k^2$ is proportional to the scalar curvature in $\mbox{AdS}_5$ spacetime; and, its scale is usually set to be $k\sim l_{P}^{-1}$, the scale of inverse Planck length. The effective potential $V(z)$, then, becomes negative for that the Compton wave length of the particle $\lambdabar\equiv\frac{\hbar}{m_0c}$ is larger than the twice Planck length $2l_{P}$; that is, for $m_0\lesssim \frac{1}{2}M_{P}=\frac{\hbar k}{2c}$.

A particular interesting choice of the proper mass of the particle is to put $m_0=\frac{1}{2}\frac{\hbar k}{c}=\frac{1}{2}M_P$, to which the effective potential $V(z)$ vanishes. In this case, Eq.(\ref{master-wave}) is reduced to the type of Eq.(\ref{KG-function}) in section 2. This is a result of fine tuning of the mass $m_0$ in UV brane, but the form of (\ref{Green-1}) approximately holds even for the particle with $m_0c^2\lesssim$ GUT, since, then, $\Delta/z^2\ll 1$ in very wide region of $z$. In what follows, we confine our attention to this particular interesting case only.

\subsection{Difference type of equation for a tuned-mass particle}

The Klein-Gordon type of fields $\{\tilde{\Psi}\}$ are wave functions in $\Delta=0$ system; we apply the boundary condition  $\tilde{\Psi}(x,1)=\tilde{\Psi}(x,L_z)=0$, by which the independent basis in $z$ space are given by $\phi^{S}_n(z)=\sqrt{\frac{2}{L_{(z)}}}\sin\left(\frac{\pi n(z-1)}{L_{(z)}}\right),\,(n=1,2,\cdots)$ with $L_{(z)}=L_z-1$. Then, the Green function $(\hat{\mathcal{K}}^{-1}_{[\beta]})_{ba}$ can be identified with $G_{ba}$ in Eq.(\ref{Green-1}) under the substitution $(y,m,L)=(k^{-1}z,0,L_{(z)})$. 

Now, inverting the equation $\hat{\mathcal{K}}_{[\beta]}\tilde{\Psi}=\tilde{\Phi}$ so that $\tilde{\Psi}=\hat{\mathcal{K}}^{-1}_{[\beta]}\tilde{\Phi}$; further, remembering $\tilde{\Psi}=\sqrt{z}\Psi^\prime$ and $\tilde{\Phi}=\sqrt{z}\Phi^\prime$, one can write $\Psi^\prime =z^{-1/2}\hat{\mathcal{K}}_{[\beta]}^{-1}z^{1/2}\Phi^\prime$. This equation suggests to use $\bm{\Psi}=U^{-1}(\sqrt{z}\Psi^\prime)$ and $\bm{\Phi}=U^{-1}(\sqrt{z}\Phi^\prime)$ instead of $\Psi=U^{-1}\Psi^\prime$ and $\Phi=U^{-1}\Phi^\prime$; then, we arive at the equation
\begin{align}
 \bm{\Psi}(x_b,z_b) =\int d^4x_a\int dz_aG_{ba}(\bar{x}_{ba},\bar{z}_{ba}-k\beta\bar{x}^0_{ba},\breve{z}_{ba}-k\beta\breve{x}^0_{ba})\bm{\Phi}(x_a,z_a), \label{a-to-b (z)}
\end{align}
where $\bm{\Psi}(x,z)=\sqrt{z-k\beta x^0}\Psi(x,z)$ and the same is ture for $\bm{\Phi}(x,z)$. The last step to make reduce Eq.(\ref{a-to-b (z)}) to one in IR brane, we  put $z_a=z_b=L_z$. Then writing $\bm{\Phi}(x_a,z_a)=\delta(z_a-L_z)\bm{\Phi}_a(x_a)$, we obtain
\begin{align}
 \bm{\Psi}_b[\bm{\Phi}_a,x_b] =\bm{\Psi}(x_b,L_z)=\int d^4x_a\mathcal{G}_{ba}\bm{\Phi}_a(x_a),
\end{align}
where 
\begin{align}
 \mathcal{G}_{ba}(\bar{x}_{ba},\breve{x}^0_{ba})=G_{ba}(\bar{x}_{ba},-k\beta\bar{x}^0_{ba}, L_z -k\beta\breve{x}^0_{ba}).
\end{align} 

Since the Green function $G_{ba}$ have the periods $2L_{(z)}$ and $L_{(z)}$ with respect to the second and the third arguments respectively; and so, the restricted green function $\mathcal{G}_{ba}(\bar{x}_{ba},\breve{x}^0_{ba})$ has the period $L_e=2L_{(z)}/k\beta$ with respect to $\breve{x}^0_{ba}$. Thus one can derive, again, the finite-time displacement equation
\begin{align}
e^{\pm L_e(\partial_0)_b}\bm{\Psi}_b[\bm{\Phi}_a,x_b] &= \int d^4x_a\left\{e^{\mp L_e(\partial_0)_a}\mathcal{G}_{ba}(\bar{x}_{ba},\breve{x}^0_{ba} \pm L_e)\right\}\bm{\Phi}_a(x_a)  \nonumber \\
 &=\int d^4x_a\mathcal{G}_{ba}(\bar{x}_{ba},\breve{x}^0_{ba})\bm{\Phi}_a(x^0_a\pm L_e, x^i_a)., \label{displacement-k}
\end{align}
this should be compared with Eq.(\ref{displacement}). The $\bm{\Psi}_b[\bm{\Phi}_a,x_b]$ is a functional of $\bm{\Phi}_a$ dressed by the effect of extra dimension under the condition $\Delta=0$. If we suppose $\bm{\Phi}_a$ to be a form such as (\ref{source}), then the $\bm{\Psi}_b$ will satisfy the domain type of field equation (\ref{domain-1}). In spite of this similarity between Eq.(\ref{displacement}) and Eq.(\ref{displacement-k}), it should be noticed that the field characterized by Eqs.(\ref{K-beta-2}) and (\ref{u-constraint-2}) is not $\bm{\Psi}_b[\bm{\Phi}_a,x_b]$ but $\Psi(x_b,L_z)=(L_z-k\beta x^0_b)^{-\frac{1}{2}}\bm{\Psi}_b[\bm{\Phi}_a,x_b]$. This means that the domain type of field equation written by $\Psi(x_b,L_z)$ has an a little complicated form than Eq.(\ref{domain-1}) reflecting the $k\neq 0$, curved spacetime effect, even for $\Delta=0$.

\section{A possible model of spinning particle in $\mbox{AdS}_5$ spacetime}

Until now, the time dependence of extra dimension has been introduced without regard to the Lorentz invariance of the Lagrangian through the substitutions such as $dy\rightarrow dy-\beta e^{-ky}dx^0$. A simple way to recover the Lorentz invariance is to replace $dx^0$ by $V\cdot dx$, where $V=(V^\mu,0)$ is a timelike Lorentz vector constructed out of the dynamical variables inherent in the particles under consideration. As a model of such a particle, we study a spinning particle characterized by the world-line interval such that
\begin{align}
\begin{split}
ds_V^2 &=g_{\hat{\mu}\hat{\nu}}\delta x^{\hat{\mu}}\delta x^{\hat{\nu}}~~\left(\, \delta x^{\hat{\mu}}=(e^{\prime\hat{\mu}}\cdot\dot{x})d\tau \, \right), \\
 (e^{\prime\,\hat{\mu}}) &=(e^{\mu}, e^5 -\beta e^{-ky}V) 
\end{split}  \label{ds-V}
\end{align}
where $g_{\hat{\mu}\hat{\nu}}$ is the RS spacetime metric; and, we read $(e^{\prime\,\hat{\mu}}\cdot\dot{x})$ as an effective velocity of the particle relative to the background spacetime. 

Now, we construct the four vector $(V^\mu)$ in each $y$-fixed brane by means of the bilinear representation of spinors so that $V^\mu \equiv \bar{\zeta}\gamma^\mu \theta,\,(\bar{\zeta}=\zeta^T\gamma^0)$, where $\theta^A=(\theta^A)^*$ and $\zeta^A=(\zeta^A)^*\,,\,(A=1,2,3,4)$ are Majorana spinors, and the $\gamma^\mu,\,(\mu=0,1,2,3)$ are $\gamma$-matrices in Majorana representation. Here, the $(\theta,\zeta)$ may be either Grassmann variables or ordinary variables; in what follows, we discuss the former case only for simplicity.

The Lagrangian for the particle in this model is set as
\begin{align}
 \mathcal{L}=\frac{1}{2}\left\{\frac{1}{e}\left(\frac{ds_{\tiny V}}{d\tau}\right)^2-(\kappa c)^2 e \right\}+i\hbar\bar{\zeta}^A\frac{d\theta^A}{d\tau}\,, \label{spinning-particle}
\end{align}
where $-i\hbar\bar{\zeta}^A$ is the momentum conjugate to $\theta^A$. We put simply $\kappa$ as a constant, although it may be a scalar function constructed out of $(\theta,\zeta)$ in general. In q-number theory, thus, we require the anti-commutator $\{\bar{\zeta}^A,\theta^B\}=\delta^{AB}$, which is equivalent to $\{\theta^A,\zeta^B\}=\{\zeta^A,\theta^B\}=-(\gamma^0)^{AB}$; more details on the spin representations in the present model will be given in Appendix B.

Varying the Lagrangian (\ref{spinning-particle}) with respect to $e$, one can derive one constraint corresponding to a wave equation in q-number theory. According to the prescription in subsection\,\ref{Tuning}, we deal with this constraint, by introducing an auxiliary canonical pair $(u,p_u)$, in the form of a pair of constrains 
\begin{align}
 \mathcal{K}_S &\equiv e^{2ky}\left(p-V p_u \right)^2+\left\{p_y^2+(\kappa c)^2\right\}=0, \\
 \phi_S &\equiv p_u+\beta e^{-ky}p_y=0.
\end{align}
Under the decomposition $V^\mu=V_{\|}^\mu+V_\perp^\mu$, where $V_{\|}^\mu=-\underline{p}^\mu(\underline{p}\cdot V),\,V_\perp^2=V^2-V_{\|}^2$, and $\underline{p}^\mu=p^\mu/\sqrt{-p^2},\,(\underline{p}^2=-1)$, the $\mathcal{K}_S$ can also be written as
\begin{align}
 \mathcal{K}_S=e^{2ky}(p-V_{\|}p_u)^2+V_\perp^2\beta^2p_y^2+\left\{p_y^2+(\kappa c)^2\right\}.
\end{align}
We, here, regard that the $V_\perp^2=V^2-V_{\|}^2$ takes a constant associated with the representation of $(\theta, \zeta)$ in q-number theory. In other words, the explicit form of $\mathcal{K}_S$ is determined for respective spin representations.

Now, similarly in subsection\,\ref{Tuning}, we define the operator $\hat{\mathcal{K}}_{(S)}=e^{-2ky}\hat{\mathcal{K}}_S$ in the q-number theory such that
\begin{align}
 \hat{\mathcal{K}}_{(S)}=\left(\hat{p}-V_{\|}\hat{p}_u\right)^2+e^{-2ky}\left\{\left(V_\perp^2\beta^2+1\right)\hat{p}_y^2+(\kappa c)^2\right\},
\end{align}
and $\hat{\mathcal{K}}_S\Psi=0$ is regarded as one-particle wave equation
\footnote{
The $(\theta,\zeta)$ and $V_\mu=\bar{\zeta}\gamma_\mu\theta$ are spinors and a four vector on the functional space of $\{\Psi\}$, since $\{x^\mu\}$ are coordinates of $M_4$ in each $y$-fixed brane of RS spacetime. In this sense, for example,  $\Psi^*V_\mu\Psi$ is a vector field in $M_4$
}
 intrinsic to this spinning particle. Then, with consideration for $[(x\cdot V_{\|}), \underline{\hat{p}}^\mu]=[(x\cdot V_{\|}), V_{\|}^\mu]=0$ and $[(x\cdot V_{\|}), \hat{p}^\mu]=i\hbar V_{\|}^\mu$, one can remove the $V_{\|}\hat{p}_u$ in $\hat{\mathcal{K}}_{(S)}$ by the transformation $\hat{\mathcal{K}}_{(S)}^\prime=U\hat{\mathcal{K}}_{(S)}U^{-1}$ with $U=e^{-\frac{i}{\hbar}(x\cdot V_{\|})\hat{p}_u}$; and, the $\hat{\phi}_S$ stays unchanged under this transformation so that $\hat{\phi}_S^\prime=U\hat{\phi}_SU^{-1}=\hat{\phi}_S$. Thus using the variable $z=e^{ky}$ as usual, the operators $\hat{\mathcal{K}}_{(S)}^\prime$ and $\hat{\phi}_S^\prime$ become
\begin{align}
 \hat{\mathcal{K}}_{(S)}^\prime &=\hat{p}^2+A\left\{\left(\hat{p}_z-i\frac{\hbar k}{2z}\right)^2 -(\hbar k)^2\frac{\Delta^\prime}{z^2}\right\}, \\
 \hat{\phi}_S^\prime &=\hat{p}_u+\beta\hat{p}_z,
\end{align}
where $A=V_\perp^2\beta^2+1$ and
\begin{align}
 \Delta^\prime=\left(\frac{1}{2}\right)^2-\left(\frac{\kappa c}{\sqrt{A}\hbar k}\right)^2.
\end{align}
In this stage, we eliminate $\hat{\phi}^\prime_S$ by reading $\hat{p}_u\equiv -\beta\hat{p}_z$ as the definition of $\hat{p}_u$; after that, applying the transformation $\hat{\mathcal{K}}_{[S]}=z^{1/2}\hat{\mathcal{K}}_{(S)}^\prime z^{-1/2}$ , we finally arrive at the expression
\begin{align}
\hat{\mathcal{K}}_{[S]}=\hat{p}^2+A\left\{p_z^2 -(\hbar k)^2\frac{\Delta^\prime}{z^2}\right\}.
\end{align}
If we transcribe the variable for $z^\prime=z/\sqrt{A}$, then the field equation $\hat{\mathcal{K}}_{[S]}\tilde{\Psi}=0$ will coincide with $\hat{\mathcal{K}}_{[\beta]}\tilde{\Psi}=0$ in section 3. Hereafter, we again consider the case $\Delta^\prime=0$ realized for $\kappa/\sqrt{A}=\frac{1}{2}M_P$, where as shown in Appendix B, the operator $A$ takes the eigenvalues $1,\,3\beta^2+1,\cdots$ respectively in the spin $0,\frac{1}{2},\cdots$ representation spaces.

The Green function $(\hat{\mathcal{K}}_{[S]}^{-1})_{ba}$ in $\{\tilde{\Psi}\}$ space, has the same form as $G_{ba}$ in Eq.(\ref{Green-1}) by reading $(y,m,L)$ as $(z^\prime=z/\sqrt{A},\kappa,L_{(z)}/\sqrt{A})$; that is, in $z$ representation, one can write $(\hat{\mathcal{K}}_{(S)}^{-1})_{ba}=G_{ba}(\bar{x}_{ba},\bar{z}_{ba}/\sqrt{A},\breve{z}_{ba}/\sqrt{A})$. Then, following the same path discussed in subsection\,\ref{Tuning}, one can solve $\mathcal{K}_{(S)}\Psi=\Phi$ in the form $\Psi=U^{-1}(z^{-1/2}\hat{\mathcal{K}}^{-1}_{[S]}z^{1/2})U\Phi$; that is, we obtain
\begin{align}
 \bm{\Psi}(x_b,z_b) &=\frac{1}{\sqrt{A}}\int d^4x_a\int dz_a G_{ba}\left(\bar{x}_{ba},\frac{\bar{z}_{ba}-k\beta (\bar{x}_{ba}\cdot V_{\|})}{\sqrt{A}},\frac{\breve{z}_{ba}-k\beta (\breve{x}_{ba}\cdot V_{\|})}{\sqrt{A}}\right) \nonumber \\
 &\times \bm{\Phi}(x_a,z_a), \label{a-to-b-spin}
\end{align}
where $\bm{\Psi}(x,z)=\sqrt{z-k\beta(x\cdot V_{\|})}\Psi(x,z)$ and the same is true for $\bm{\Phi}(x,z)$. Finally, setting $z_b=z_a=L_z$ and $\bm{\Phi}(x_a,z_a)=\bm{\Phi}_a(x_a)\delta(z_a-L_z)$, we obtain the expression of Eq.(\ref{a-to-b-spin}) on the IR brane so that
\begin{align}
 \bm{\Psi}_b[\bm{\Phi}_a,x_b]\equiv\bm{\Psi}(x_b,L_z)=\int d^4x_a\mathcal{G}^{(S)}_{ba}\bm{\Phi}_a(x_a),
\end{align}
where
\begin{align}
 \mathcal{G}^{(S)}_{ba}=\frac{1}{\sqrt{A}}G_{ba}\left(\bar{x}_{ba},\frac{-k\beta (\bar{x}_{ba}\cdot V_{\|})}{\sqrt{A}},\frac{L_z-k\beta (\breve{x}_{ba}\cdot V_{\|})_b}{\sqrt{A}}\right). \label{a-to-b(S)}
\end{align}

Now, remembering that the Green function $G_{ba}$ in Eq.(\ref{a-to-b(S)}) has the periods $2L_{(z)}/\sqrt{A}$ and $L_{(z)}/\sqrt{A}$ with respect to the second and the third arguments respectively, one can derive the finite displacement equation
\begin{align}
  e^{\pm L_e(V_{\|}\cdot\partial)_b}\bm{\Psi}_b[\bm{\Phi}_a,x_b] &=\int d^4x_a\mathcal{G}^{(S)}_{ba}\bm{\Phi}_a(x_a \pm L_e V_{\|}), \label{displacement-S}
\end{align}
where $L_e=2L_{(z)}/(k\beta V_{\|}^2)$, and the $V_{\|}^2$ has been treated as a number given by the spin representation under consideration. The discussion after this, thus, can be done in parallel with section 2 and section 3. For example, for (+) sign of Eq.(\ref{displacement-S}), let us consider the case such that
\begin{align}
 \bm{\Phi}_a(x)=\theta(\hat{p}^0)\delta\left(\hat{p}^2+(mc)^2\right)\phi_a(x),~\left(\, m=e^{-kL}\kappa/\sqrt{A} \,\right),
\end{align}
in order to regard $\bm{\Psi}_b$ as a domain type of free field, a dressed functional of $\bm{\Phi}_a$, then with the aid of $V_{\|}\cdot \partial=V\cdot \partial$, we arrive at the equation 
\begin{align}
 e^{L_e(V\cdot\partial)_b}\bm{\Psi}_b[\bm{\Phi},x_b]=e^{\frac{i}{\hbar}L_e\left(-V^0\sqrt{\hat{\bm{p}}^2+(mc)^2}+\bm{V}\cdot\hat{\bm{p}}\right)_b}\bm{\Psi}_b[\bm{\Phi}_a,x_b]. 
\end{align}
This is an example of the domain type of field equation invariant under the Lorentz transformation. This is the result that is wanted to derive based on the particle models embedded in $\mbox{AdS}_5$ spacetime.

\section{Summary and discussion}

In this paper, we have studied the possibility to regard Yukawa's domain type of field as an effective field associated with the particles embedded in the $\mbox{AdS}_5$ spacetime with a warp factor $e^{-2ky}$. The keys to get such an effective theory are threefold; the first is a periodic structure of Green function of the particle in the $\mbox{AdS}_5$ spacetime with respect to the fifth coordinate variable $y$; the second is the mixing between the time and the fifth directions in some way, the third is the choice of the function space of initial source fields in IR brane. 

In more detail, the fifth dimension in the $\mbox{AdS}_5$ spacetime with the warp factor is the one dimensional orbifold, in which the particles suffer an infinite square-well potential under the Dirichlet type of boundary conditions at both ends of the fifth dimension. Then there arises a periodic structure to the Green function just like the problem of the particle in a box. 

When $x^0$ and $x^5(=y)$ are mixed in the Green function, its periodic structure with respect to the $x^5$ variable will be copied to the $x^0$ variable. In a sense, the mixing is an introduction of a time-dependent extra dimension or a moving-extra dimension for the particle. When we write the world-line interval of a particle with of the background spacetime  metric $g_{\hat{\mu}\hat{\nu}}$ so that $ds^2=g_{\hat{\mu}\hat{\nu}}\delta x^{\hat{\mu}}\delta x^{\hat{\nu}},\,(\, \delta x^{\hat{\mu}}=(e^{\prime\,\hat{\mu}}\cdot\dot{x})d\tau$, the mixing is  attributed to the velocity $(e^{\prime\,\hat{\mu}}\cdot\dot{x})$.

To make clear the mixing, we first studied two types of toy models of particles in $M_4\otimes S^1/Z_2~(k=0)$, to which the velocity $(e^{\prime\,\hat{\mu}}\cdot\dot{x})$ represents that the particles are placed in the spacetime with a boosted or a rotated extra dimension. In those models, we could derive in IR brane a finite-time displacement equation for the field $\Psi_b$ generated by the propagation of a source field $\Phi_a$; here, we call the hyper-surface at largest $y$ end as IR brane even in $M_4\otimes S^1/Z_2$ models. 

Next, we extended those models to one in AdS${}_5$ spacetime $(k\neq 0)$. Though the Green function becomes complex in this case, the specific choice of proper-mass of the particle in UV brane such as $m_0=\frac{1}{2}M_P$ make reduce the problem to one in $M_4\otimes S^1/Z_2$ type of spacetime. With such a fine tuning of parameters, a time-displacement equation for the field $\Psi_b$ out of a source field $\Phi_a$ is obtained in this model too. In those models, however, the Lorentz invariance of the formalism on each y-fixed brane is lost, since the mixing of variables is set between $x^0$ and $x^5$. 

So, we lastly studied a model of spinning particle, in which the mixing of variables is caused between $V\cdot\dot{x}$ and $x^5$, where $V^{\mu}$ is a four vector living in each brane, and is constructed out of the spin degrees of freedom of the particle. As a result, the Lorentz invariance is restored in this model; and, a finite-time displacement equation again holds for the pair of $\Psi_b$ and $\Phi_a$ under a fine tuning of $\kappa$ similar to the previous models. 

From those results, one can say the following: 1) The fine tuning $m_0=\frac{1}{2}M_P$ gives rise to the mass of the particle in IR brane such as $m=m_0e^{-kL}\,(\mbox{e.g.}~\lesssim 3.7\,\mbox{TeV}/c^2~\mbox{for}~kL\gtrsim 35)$, which implies the existence of a specific mass scale related to an interesting physics in IR brane.  2) When we write the relation between $\Psi_b(x_b)$ and $\Phi_a(x_a)$ in IR brane as $\Psi_b[\Phi_a,x_b]$, one can read that the $\Psi_b[\Phi_a,x_b]$ is a dressed functional of $\Phi_a$ and a function of $x_b$ in the effect of extra dimension. The importance is that the finite-time displacement equation for the pair of $\Psi_b$ and $\Phi_a$ can be rewritten as a domain type of difference equation for $\Psi_b[\Phi_a,x_b]$ provided that the source field $\Phi_a$ belongs to a suitable functional space $\{\Phi^S\}$ characterized, for example, by Eq.(\ref{suitable source}). Since this functional space is rather wide one including the free on-mass-shell fields, Yukwa's domain type of field is not so extraordinary from the viewpoint of the effective field $\Psi_b[\Phi^S_a,x_b]$, which can be represented symbolically as
\begin{align}
 \left(\Psi_b\right)_{\mbox{Domain}} = \int d^4x_a\Big(\mathcal{G}_{ba}\Big)_{\mbox{AdS boundary}}~\Phi_a^S~. \label{domain-field}
\end{align}
Here, the elementary local field is not $\Psi_b$ but $\Phi_a^S$; this is the difference between the present formalism and that of Yukawa, which treats $\Psi_b$ as an elementary non-local field. 

We also note that the Green function $\mathcal{G}_{ba}=G_{ba}\big|_{\mbox{boundary}}$ defines a specific graph in the product space $\Psi_b \times \Phi_a$; and, it will be significant attempt as the next task to attack the AdS spacetime structure associated with $G_{ba}$ conversely from the viewpoint of the graph structure.

Finally, we comment the following:  in a preceding article\,\cite{kappa-M}, we showed that the $\kappa$-Minkowski structure based on the AdS${}_{5}$ spacetime can derive a domain type of field equation in $M_4$. The field equations in such a non-commutative spacetime include the terms $e^{\pm\lambda(\partial_X)^2}\Psi$ instead of the directional difference terms $e^{\pm \lambda^\mu(\partial_X)_\mu}\Psi$. Since, however, the $\kappa$-Minkowski type of field equations are realized by linear combinations of directional difference type of field equations, it is also interesting to study the relation between two types of field equations. Another emphatic point is that if we consider the presence of gauge fields in IR brane, then the gauge fields can play a similar role of vector $V^\mu$ in spinning particles; that is, the Lorentz invariant formalism will be realized in this case too. Those are the interesting forthcoming problems.

\section*{Acknowledgments}

The authors wish to thank the members of the theoretical group in Nihon University for their hospitality and interest in this work.

\appendix

\section{On a modified RS spacetime} \label{spacetime}

In association with the discussion in section 3, we here study the spacetime models characterized by the line element
\begin{align}
 ds^2 &=e^{-2kx^5}\left\{-(Adx^0+Bdx^5)^2+dx^idx^i\right\}+\left\{Cdx^5+D(x^5)dx^0\right\}^2 \nonumber \\
 &=\eta_{\hat{\mu}\hat{\nu}}e^{(\hat{\mu})}{}_{\hat{\alpha}} e^{(\hat{\nu})}{}_{\hat{\beta}}dx^{\hat{\alpha}}dx^{\hat{\beta}} \\
 &~ \left(D(x^5)=D_0e^{-kx^5};\,A,B,C,D_0=\mbox{const.}\right), \nonumber
\end{align}
where $e^{(0)}{}_{\hat{\mu}}=e^{-kx^5}(A\delta^0_{\hat{\mu}}+B\delta^5_{\hat{\mu}})$, $e^{(i)}{}_{\hat{\mu}}=e^{-kx^5}\delta^i_{\hat{\mu}}$, and $e^{(5)}{}_{\hat{\mu}}=C\delta^5_{\hat{\mu}}+D(x^5)\delta^0_{\hat{\mu}}$; further,  the world metric $g_{\hat{\mu}\hat{\nu}}=\eta_{\hat{\alpha}\hat{\beta}}e^{(\hat{\alpha})}{}_{\hat{\mu}}e^{(\hat{\beta})}{}_{\hat{\nu}}$ is given by
\begin{align}
(g_{\hat{\mu}\hat{\nu}}) =
\begin{bmatrix}
-e^{-2kx^5}A^2+\mathcal{D}^2 & 0 & 0 & 0 & -e^{-2kx^5}AB+CD \\
0 & e^{-2kx^5} & 0 & 0 & 0 \\
0 & 0 & e^{-2kx^5} & 0 & 0 \\
0 & 0 & 0 & e^{-2kx^5} & 0 \\
-e^{-2kx^5}AB+CD & 0 & 0 & 0 & -e^{-2kx^5}B^2+C^2
\end{bmatrix}. \label{metric}
\end{align}

Now, using the 1-forms $\{\bm{d}x^{\hat{\mu}}\}$, the exterior derivatives of $x^{\hat{\mu}}$'s characterized by $\bm{d}x^{\hat{\mu}}\bm{d}x^{\hat{\nu}}=-\bm{d}x^{\hat{\nu}}\bm{d}x^{\hat{\mu}}$  and $\bm{d}(\bm{d}x^{\hat{\mu}})=0$, let us define the 1-forms in the local-Lorentz basis such as $\bm{\omega}^{(\hat{\mu})}=e^{(\hat{\mu})}{}_{\hat{\nu}}\bm{d}x^{\hat{\nu}}$, to which the inverted relations $\bm{d}x^{\hat{\mu}}=e^{\hat{\mu}}{}_{(\hat{\nu})}\bm{\omega}^{(\hat{\nu})},\,(e^{\hat{\mu}}{}_{(\hat{\nu})}e^{(\hat{\nu})}{}_{\hat{\rho}}=\delta^{\hat{\mu}}{}_{\hat{\rho}})$ become 
\begin{align}
 \bm{d}x^0 &=M^{-1}\left\{Ce^{kx^5}\bm{\omega}^{(0)}-B\bm{\omega}^{(5)}\right\}, \\
 \bm{d}x^i &=e^{kx^5}\bm{\omega}^{(i)}, \\
 \bm{d}x^5 &=M^{-1}\left\{-D_0\bm{\omega}^{(0)}+A\bm{\omega}^{(5)}\right\} ,
\end{align}
where $M=AC-BD$. Then, it is not difficult to find the connection forms defined by $\bm{d}\bm{\omega}^{(\hat{\mu})}=-\bm{\omega}^{(\hat{\mu})}{}_{(\hat{\nu})}\wedge\bm{\omega}^{(\hat{\nu})},\,(\bm{d}^2=0)$ so as to be
\begin{align}
\left(\bm{\omega}^{(\hat{\mu})}{}_{(\hat{\nu})}\right)=M^{-1}kA
\begin{bmatrix}
 0 & A^{-1}D_0\bm{\omega}^{(j)} & -\bm{\omega}^{(0)}+A^{-1}D_0\omega^{(5)} \\
 A^{-1}D_0\bm{\omega}^{(i)}  & 0 & -\bm{\omega}^{(i)} \\
-\bm{\omega}^{(0)}+A^{-1}D_0 \bm{\omega}^{(5)} & \bm{\omega}^{(j)} & 0
\end{bmatrix}. \label{connection}
\end{align} 

In terms of those connection forms, the curvature 2-forms are defined by $\bm{\mathcal{R}}^{(\hat{\mu})}{}_{(\hat{\nu})}=d\bm{\omega}^{(\hat{\mu})}{}_{(\hat{\nu})}+\bm{\omega}^{(\hat{\mu})}{}_{(\hat{\rho})}\wedge \bm{\omega}^{(\hat{\rho})}{}_{(\hat{\nu})}$, which can be derived from (\ref{connection}) after tedious calculation in the following form:
\begin{align}
 \bm{\mathcal{R}}^{(0)}{}_{(0)} &= \bm{\mathcal{R}}^{(5)}{}_{(5)}=0 ,\\
 \bm{\mathcal{R}}^{(0)}{}_{(j)} &=(M^{-1}k)^2M^{-1}A\left[(CD_0^2-MA)\bm{\omega}^{(0)}\wedge\bm{\omega}^{(j)}+D_0BD\bm{\omega}^{(j)}\wedge\bm{\omega}^{(5)}\right] ,\\
 \bm{\mathcal{R}}^{(0)}{}_{(5)} &=(M^{-1}k)^2M^{-1}AC(D_0^2-A^2)\bm{\omega}^{(0)}\wedge\bm{\omega}^{(5)} ,\\
 \bm{\mathcal{R}}^{(i)}{}_{(j)} &=(M^{-1}k)^2(D_0^2-A^2)\bm{\omega}^{(i)}\wedge\bm{\omega}^{(j)} ,\\
 \bm{\mathcal{R}}^{(i)}{}_{(5)} &=-(M^{-1}k)^2M^{-1}ABD_0D\bm{\omega}^{(0)}\wedge\bm{\omega}^{(i)} \nonumber \\
 &~~+(M^{-1}k)^2M^{-1}(D_0^2M-A^3C)\bm{\omega}^{(i)}\wedge\bm{\omega}^{(5)} .
\end{align}
Introducing, here, the tangent basis $\bm{E}_{(\hat{\mu})}=e_{(\hat{\mu})}{}^{\hat{\rho}}\partial_{\hat{\rho}}$, to which the inner products $\bm{\omega}^{(\hat{\mu})}\cdot\bm{E}_{(\hat{\nu})}=\delta^{\hat{\mu}}{}_{\hat{\nu}}$ and $(\bm{\omega}^{(\hat{\mu})}\wedge\bm{\omega}^{(\hat{\nu})}): \bm{E}_{(\hat{\alpha})}\bm{E}_{(\hat{\beta})}=(\delta^{\hat{\mu}}_{\hat{\alpha}}\delta^{\hat{\nu}}_{\hat{\beta}}-\delta^{\hat{\nu}}_{\hat{\alpha}}\delta^{\hat{\mu}}_{\hat{\beta}})$ hold, one can write the cuarvature tensor as $R^{(\hat{\rho})}{}_{(\hat{\mu})(\hat{\sigma})(\hat{\nu})}=\bm{\mathcal{R}}^{(\hat{\rho})}{}_{(\hat{\mu})}:\bm{E}_{(\hat{\sigma})}\bm{E}_{(\hat{\nu})}$, from which the Ricci tensor $R_{(\hat{\nu})(\hat{\beta})}= R^{(\hat{\mu})}{}_{(\hat{\nu})(\hat{\mu})(\hat{\beta})}$ in local Lorentz basis becomes
\begin{align}
 R_{(0)(0)} &=-(M^{-1}k)^2M^{-1}\left[4AC(D_0^2-A^2)+3BDA^2\right] ,\\
 R_{(0)(5)} &=(M^{-1}k)^2M^{-1}3BDAD_0 ,\\ 
  R_{(i)(j)} &=(M^{-1}k)^2M^{-1}\left[4AC(D_0^2-A^2)-3BD(D_0^2-A^2)\right]\delta_{ij} ,\\
 R_{(5)(5)} &=(M^{-1}k)^2M^{-1}\left[4AC(D_0^2-A^2)-3BDD_0^2\right] .
\end{align}
in addition to $R_{(0)(i)}=R_{(i)(5)}=0$. The results are summarized as follows
\begin{align}
\begin{split}
 R_{(\hat{\mu})(\hat{\nu})} &=(M^{-1}k)^2M^{-1}\Big[\left\{4AC(D_0^2-A^2)+\delta_{(\hat{\mu})}\right\}\eta_{\hat{\mu}\hat{\nu}}  \\
 & \hspace{32mm} +3BDAD_0\delta^0_{\{\hat{\mu}}\delta^5_{\hat{\nu}\}}~\Big] ,
\end{split}
\end{align}
where $\delta_{(0)}=3BDA^2, \delta_{(i)}=-3BD(D_0^2-A^2)$, and $\delta_{(5)}=-3BDD_0^2$; those results give rise to the scalar curvature such that
\begin{align}
 R &=\eta^{\hat{\mu}\hat{\nu}} R_{(\hat{\mu})(\hat{\nu})}=(M^{-1}k)^2M^{-1}(20AC-12BD)(D_0^2-A^2) .
\end{align}

Now, the RS spacetime is the specific case with $A=C=1$ and $B=D=0$; then, the scalar curvature becomes the well-known result $R=-20k^2$. Further, for the spacetime with $A=C=1$, $D_0=-\beta$, and $B=0$, which is corresponding to the model of spacetime with the rotating $S^1/Z_2$ extra dimension, the scalar curvature $R=-20k^2(1-\beta^2)$ is obtained. Roughly speaking, in the spacetime with $B=0$, the Ricci tensor in local Lorentz basis is different from that of RS spacetime only by the scale of $k$; and, the spacetime has the constant scalar curvature $R=20k^2(D_0^2-A^2)/(AC)^2$. In other words, the $B=0$ case becomes a solution of the Einstein equation under a similar energy-momentum tensor to one in RS spacetime.

Finally, it should be pointed out that the particular case of $D_0^2-A^2=0$ leads to the spacetime with vanishing scalar curvature; if we require simultaneously $B=0$, then the Ricci tensor in this spacetime will vanish; that is, the spacetime is a Ricci flat one being no longer AdS${}_5$.

\section{Spin representation space}

In this paper, the gamma matrices characterized by $\{\gamma_\mu,\gamma_\nu\}=2\eta_{\mu\nu}$ are taken to be the Majorana representation, which are given explicitly by
\begin{align}
 \gamma^0 =i\rho_2\otimes\sigma_1,~\gamma^1=\rho_1\otimes\sigma_0,~\gamma^2=\rho_2\otimes\sigma_2,~\gamma^3=\rho_3\otimes\sigma_0,
\end{align}
where $\rho_i$ and $\sigma_i\,(i=1,2,3)$ are two sets of Pauli matrices, and $\sigma_0$ is $2\times 2$ unit matrix. Since the $\gamma^\mu$'s are real component matrices, we may deal with $(\theta,\zeta)$ as real four-component spinors; then, $(\bar{\theta},\bar{\zeta})=(\theta^T\gamma^0,\zeta^T\gamma^0)$. Within the framework of pseudo-classical mechanics\cite{Berezin, Casalbuoni} (classical mechanics\cite{Supple}), the spinors can be treated as fermionic (bosonic) dynamical variables of spinning particles. It is also obvious that $\zeta^C=\gamma^0\bar{\zeta}^T=\zeta$ and $\theta^C=\theta$ for real four spinors; that is, the charge conjugation of $\zeta$ and $\theta$ are themselves. 

The Lagrangian (\ref{spinning-particle}) says that the $-i\hbar\bar{\zeta}^A\,\left(=\frac{\partial}{\partial\dot{\theta}^A}\mathcal{L}\,\right)$ and $\theta^A,\,(A=1,2,3,4)$ are canonical pairs; and, since we confine to the case of fermionic spinors, we require in q-number theory the anti-commutation rules
\begin{align}
 \{\theta^A,\bar{\zeta}^B \} \equiv \theta^A\bar{\zeta}^B+\bar{\zeta}^B\theta^A=\delta^{AB},~~\{\theta^A,\theta^B\}=\{\bar{\zeta}^A,\bar{\zeta}^B\}=0.  \label{anti-comm}
\end{align}
 The generators of Lorentz transformation for the intrinsic spins become
\begin{align}
 S_{\mu\nu}=\frac{1}{2}\bar{\zeta}\sigma_{\mu\nu}\theta\,,~~\left(\,\sigma_{\mu\nu}=\frac{i}{2}[\gamma_\mu,\gamma_\nu]\,\right),
\end{align}
by which the infinitesimal transformations are written as
\begin{align}
 \delta\theta &=\frac{i}{2}\delta\omega^{\mu\nu}[S_{\mu\nu},\theta]=-\frac{i}{4}\delta\omega^{\mu\nu}\sigma_{\mu\nu}\theta, \\
 \delta\bar{\zeta} &=\frac{i}{2}\delta\omega^{\mu\nu}[S_{\mu\nu},\bar{\zeta}]=\frac{i}{4}\delta\omega^{\mu\nu}\bar{\zeta}\sigma_{\mu\nu}.
\end{align}
Under those transformations, one can verify that $\bar{\zeta}\theta$ and $V_\mu=\bar{\zeta}\gamma_\mu\theta$ transform respectively as scalar and vector quantities; that is, that $\delta(\bar{\zeta}\theta)=0$ and $\delta V_\mu=-\frac{1}{2}\delta\omega^{\rho\sigma}\left(V_\rho\eta_{\sigma\mu}-V_\sigma\eta_{\rho\mu}\right)$.

Now, the definition of $S_{\mu\nu}$ implies that the ground state, the spin $0$ state, and its adjoint state should be defined by $\theta|0\rangle=0,\, \langle\bar{0}|\bar{\zeta}=0$ and $\langle\bar{0}|0\rangle=1$. Then the states $|\alpha\rangle=\bar{\zeta}^\alpha|0\rangle$ and $\langle\bar{\alpha}|=\langle\bar{0}|\theta^\alpha$ form a pair of spin $1/2$ states, from which one can verify $\langle\bar{\alpha}|\beta\rangle=\delta_{\alpha\beta},\,\langle\bar{\alpha}|V_\mu|\beta\rangle=(\gamma_\mu)_{\alpha\beta}$, and $\langle\bar{\alpha}|S_{\mu\nu}|\beta\rangle=\frac{1}{2}(\sigma_{\mu\nu})_{\alpha\beta}$ and so on. Further, the bi-linear combinations $\bar{\zeta}\zeta=\zeta^T\gamma^0\zeta$ and $\bar{\zeta}\gamma^\mu\zeta=\zeta^T(\gamma^0\gamma^\mu)\zeta$ are transformed respectively as scalar and vector quantities under the Lorentz transformations; and so, the states $|S_n\rangle=(\bar{\zeta}\zeta)^n|0\rangle,\,(n=0,1,2)$ form spin 0 states in this representation. On the other hand, because of $(\gamma^0\gamma^\mu)^T=(\gamma^0\gamma^\mu)$, the $\bar{\zeta}\gamma^\mu\zeta|0\rangle=\zeta^T\gamma^0\gamma^\mu\zeta|0\rangle$ type of vector state vanishes as a result of Grassmann property of $\{\zeta^A \}$. 

Now, since the particle momentum $p$ belongs to timelike orbits for $m\neq 0$, one can obtain the expression $V_\perp^2=(\zeta\gamma^i\theta)^2$ at the rest frame $p=(p^0,0,0,0)$. With the help of this expression, it is not difficult to verify that $V_\perp^2|S_n\rangle=0,(n=0,1,2)$ and $V_\perp^2|\alpha\rangle=3|\alpha\rangle$. The definition $A=V_\perp^2\beta^2+1$, then, gives rise to $A|S_n\rangle=|S_n\rangle,(n=0,1,2)$, and $A|\alpha\rangle=(3\beta^2+1)|\alpha\rangle$ and so on. In any case, the eigenvalues of $A$ are greater than $1$, and they remain in finite numbers provided that $\theta$ and $\zeta$ are Grassmann variables.

We finally comment on the Weyl spinors $(\xi,\eta)$ related to $(\theta,\zeta)$ by
\begin{align}
 \theta=\begin{bmatrix} (\xi+\xi^*) \\ -i\sigma_3(\xi-\xi^*) \end{bmatrix}\,,~~~\zeta=\begin{bmatrix} (\eta+\eta^*) \\ -i\sigma_3(\eta-\eta^*) \end{bmatrix}. \label{Weyl-1}
\end{align}
Remembering $\frac{1}{2}(1 \mp i\gamma_5)=\frac{1}{2}\begin{pmatrix} \sigma_0 & \pm i\sigma_3 \\ \mp i\sigma_3 & \sigma_0 \end{pmatrix}$, the first of Eq.(\ref{Weyl-1}) leads to
\begin{align}
  \frac{1}{2}(1-i\gamma_5)\theta &=\begin{bmatrix}\xi \\ -i\sigma_3 \xi \end{bmatrix}~,~~ \frac{1}{2}(1+i\gamma_5)\theta=\begin{bmatrix}\xi^* \\ i\sigma_3 \xi^* \end{bmatrix}~, \label{Weyl-2}
\end{align}
and the same is true for $\zeta$. Then, using $\left\{\frac{1}{2}(1\pm i\gamma_5)\right\}^2=\frac{1}{2}(1\pm i\gamma_5)$, we obtain
\begin{align}
 \eta^T\sigma_2 \xi &=-\frac{1}{2}\bar{\zeta}\frac{1-i\gamma_5}{2}\theta,~~\eta^{*T}\sigma_2 \xi^*=\frac{1}{2}\bar{\zeta}\frac{1+i\gamma_5}{2}\theta\,,
\end{align}
from which follows
\begin{align}
 \bar{\zeta}\theta=-2\left(\eta^T\sigma_2\xi-\eta^{*T}\sigma_2\xi^* \right).
\end{align}
After a little calculation, one can also obtain
\begin{align}
 V^\mu=\bar{\zeta}\gamma^\mu\theta=2\left(\tilde{\eta}^\dag\sigma^\mu\tilde{\xi}-\tilde{\xi}^\dag\sigma^\mu\tilde{\eta}\right),
\end{align}
where $\xi=U\tilde{\xi}$ and $\eta=U\tilde{\eta}$ with $U=\frac{1}{\sqrt{2}}(\sigma_0+i\sigma_1)$. 

Further, the first two-component columns of Eq.(\ref{Weyl-2})  can be written as
\begin{align}
 \xi &=\frac{1}{2}(\sigma_0,i\sigma_3)\theta ~~\left(~ \xi^*=\frac{1}{2}(\sigma_0,-i\sigma_3)\theta~\right)
\end{align}
; and so,
\begin{align}
 \xi^T &=\frac{1}{2}\bar{\theta}\begin{bmatrix} -\sigma_2 \\ 
\sigma_1 \end{bmatrix}~~\left(~ \xi^{*T}=\frac{1}{2}\bar{\theta}\begin{bmatrix} \sigma_2 \\ \sigma_1 \end{bmatrix}~\right).
\end{align}
One can write down the same equations for $\zeta$ too. Then, from Eq.(\ref{anti-comm}), it is not difficult to verify $\{\eta^a,\xi^{*b}\}=\{\eta^{*a},\xi^b\}^*=0,\,(a,b=1,2)$ and
\begin{align}
 \{\eta^a,\xi^b\}=\{\eta^{*a},\xi^{*b}\}^*=\frac{1}{2}(\sigma_2)^{ab} \,. \label{anti-comm-2}
\end{align}
If we notice that the kinetic term of spinors have the form $i\hbar\bar{\zeta}\dot{\theta}=-2i\bar(\eta^T\sigma_2\dot{\xi}-\eta^{*T}\sigma_2\dot{\xi}^*)$, one can verify $\pi_\xi=\frac{\partial}{\partial\dot{\xi}}\mathcal{L}=\frac{\partial}{\partial\dot{\xi}}(i\hbar\bar{\zeta}\dot{\theta})=i\hbar 2(\eta^T\sigma_2)$ for the Lagrangian  (\ref{spinning-particle}). Thus the canonical anti-commutation relation gives $\{(\eta^T\sigma_2),\xi \}=-\frac{1}{2}\sigma_0$, which is nothing but Eq.(\ref{anti-comm-2}).

\end{document}